\documentclass[aps,prd,twocolumn,eqsecnum,showpacs,amsmath,nofootinbib]{revtex4}

\usepackage[dvips]{color,graphicx}
\usepackage{amsfonts,amssymb,theorem,mathrsfs,times}
\newcommand{\ma}[1]{\mbox{$\mathcal{#1}$}}
\newcommand{\qed}{\hbox{\rule[-2pt]{6pt}{6pt}}}
\newcommand{\D}{{\rm d}}

{\theorembodyfont{\upshape}
\newtheorem{Prop}{Proposition}}
{\theorembodyfont{\upshape}
}
{\theorembodyfont{\upshape}
\newtheorem{lm}{Lemma}}
{\theorembodyfont{\upshape}
\newtheorem{dn}{Definition}}
{\theorembodyfont{\upshape}
}
{\theorembodyfont{\upshape}
}

\newcommand{\dalm}{\kern1pt\vbox{\hrule height 0.9pt\hbox{\vrule width
0.9pt\hskip 2.5pt\vbox{\vskip 5.5pt}\hskip 3pt\vrule width 0.3pt}\hrule height
0.3pt}\kern1pt}

\begin{document}

\title{
Static and symmetric wormholes respecting energy conditions in Einstein-Gauss-Bonnet gravity
}

\author{Hideki Maeda$^{1,2}$}
\email{hideki@cecs.cl}
\author{Masato Nozawa$^3$}
\email{nozawa@gravity.phys.waseda.ac.jp}


\address{ 
$^{1}$Centro de Estudios Cient\'{\i}ficos (CECS), Arturo Prat 514, Valdivia, Chile\\
$^{2}$Department of Physics, International Christian University, 
3-10-2 Osawa, Mitaka-shi, Tokyo 181-8585, Japan\\
$^{3}$Department of Physics, 
Waseda University, Tokyo 169-8555, Japan
}

\date{\today}

\begin{abstract} 
Properties of
$n(\ge 5)$-dimensional static wormhole solutions are investigated in Einstein-Gauss-Bonnet gravity with or without a cosmological constant $\Lambda$.
We assume that the spacetime has symmetries corresponding to the
 isometries of an $(n-2)$-dimensional maximally symmetric space
with the sectional curvature $k=\pm 1, 0$.
It is also assumed that the metric is at least $C^{2}$ and the $(n-2)$-dimensional maximally symmetric subspace is compact. 
Depending on the existence or absence of the general relativistic limit $\alpha \to 0$, solutions are classified into general relativistic (GR) and non-GR branches, respectively, where $\alpha$ is the Gauss-Bonnet coupling constant. 
We show that a wormhole throat respecting the dominant energy condition coincides with a branch surface in the GR branch, otherwise the null energy condition is violated there.
In the non-GR branch, it is shown that there is no wormhole solution for $k\alpha \ge 0$.
For the matter field with zero tangential pressure, it is also shown in the non-GR branch with $k\alpha<0$ and $\Lambda \le 0$ that the dominant energy condition holds at the wormhole throat if the radius of the throat satisfies some inequality.
In the vacuum case, a fine-tuning of the coupling constants is shown to be necessary and the radius of a wormhole throat is fixed.
Explicit wormhole solutions respecting the energy conditions in the whole spacetime are obtained in the vacuum and dust cases with $k=-1$ and $\alpha>0$.
\end{abstract}

\pacs{
04.20.Gz 
04.50.-h 
04.50.Gh 
} 
\maketitle

\section{Introduction}
A wormhole is a hypothetical object in general relativity connecting two (or more) asymptotic regions or infinities.
Albeit the concept of a wormhole is global and topological, a wormhole is locally characterized by a ``throat'', which is a two-dimensional compact spatial surface of minimal area on an achronal hypersurface.
While the term {\it wormhole} was coined by Wheeler in 1957~\cite{wheeler1957}, the history of wormholes in general relativity began in 1935. 

Maximally extended Schwarzschild spacetime is the simplest example of wormhole spacetimes.
By the coordinate transformation of the radial coordinate in this spacetime, Einstein and Rosen demonstrated the first example of ``static wormholes'' in 1935, which is now called the Einstein-Rosen bridge~\cite{er1935}.
However, its wormhole throat corresponding to the bifurcation two-sphere of the event horizon is actually a coordinate singularity and not covered by those coordinates.
In this sense, the Einstein-Rosen bridge represents a static wormhole {\it without} a throat.
It is certainly possible to introduce a set of coordinates to represent the Einstein-Rosen bridge {\it with} a throat; however, as is clear in the Penrose diagram, it is realized only instantaneously, which was first pointed out by Fuller and Wheeler~\cite{fw1962}.
Thus, although the maximally extended Schwarzschild spacetime surely represents a dynamical wormhole, the Einstein-Rosen bridge is not gratifying as a static wormhole.

Although the Schwarzschild solution is a good example of dynamical wormholes satisfying energy conditions, it is not traversable, as is also clear in the Penrose diagram, where traversability is defined globally such that a (non-spacelike) observer can travel from one infinity to another.
This is attributed to the fact that there is no wormhole throat on null hypersurfaces in that spacetime. 
Traversable wormholes are quite intriguing because they admit the (apparent) superluminal travel as a global effect of the spacetime topology~\cite{visser,superluminal,lobo2007}.

In 1988, Morris and Thorne presented a metric representing a static traversable wormhole~\cite{mt1988}.
This wormhole connects two asymptotically flat spacetimes and is now a well-known classic in general relativity.
(Here it is noted that static wormhole metrics were given even before Morris and Thorne~\cite{before}.)
Although they did not specify the matter field, it was later found to be compatible with a tachyonic massless scalar field~\cite{clement1989,sh2002}.
Subsequently, Morris and Thorne discussed with Yurtsever the possibility that traversable wormholes are available to make time machines~\cite{mty1988,timemachine}.
After these seminal works, wormholes have been attracting relativists for a long time.
(The readers should refer to~\cite{visser} for a standard textbook and~\cite{lobo2007} for an excellent recent review.)

Unfortunately enough, it is well known in general relativity that exotic matter violating the null energy condition is necessary for static traversable wormholes~\cite{visser,hv1997,negative}.
In the asymptotically flat case, this is also a natural consequence of the topological censorship~\cite{TC}, which severely prohibits us from traveling to other worlds. 
It goes without saying that constructing a wormhole with ordinary matter respecting energy conditions has been a big challenge in gravitation physics.
In order to slip through the net of the strong results in general relativity, we shall consider other extended theories of gravity.
In scalar-tensor theories of gravity, for example, a non-minimally coupled scalar field could play the role of exotic matter ~\cite{bd}.
Gravity sectors in higher curvature theories of gravity could also play such a role~\cite{gb}.

Gravitation physics in higher dimensions is a prevalent subject of current research mainly motivated by string theory, and higher-dimensional Lorentzian wormhole solutions have also been investigated~\cite{kk,branewormhole,bk1992,dot2007,dot2007b}. 
In arbitrary dimensions, the most general action constructed from the Riemann curvature tensor and its contractions giving rise to the second-order quasi-linear field equations is given by the Lovelock polynomial~\cite{lovelock}. 
It surely reduces to the Einstein-Hilbert action with $\Lambda$ in four dimensions. 
Indeed, Einstein-Gauss-Bonnet gravity, whose Lagrangian includes the second-order Lovelock term as the higher curvature correction to general relativity, is achieved in the low-energy limit of heterotic string theory~\cite{Gross}. 

Bhawal and Kar considered spherically symmetric static spacetimes in Einstein-Gauss-Bonnet gravity and showed that the weak energy condition must be violated at the wormhole throat if the Gauss-Bonnet coupling constant $\alpha$ is positive.
In the case with negative $\alpha$, on the other hand, the weak energy condition holds there in the absence of the tangential pressure of the matter field~\cite{bk1992}.
Here we note that $\alpha$ is preferred to be positive from the string viewpoint~\cite{Gross}.

Recently, a vacuum static wormhole solution was obtained in five dimensions in Chern-Simons gravity, which is Einstein-Gauss-Bonnet gravity with a special combination between $\alpha$ and $\Lambda$~\cite{dot2007}.
In that solution, the spacetime is a product manifold of a two-dimensional Lorentzian manifold and a three-dimensional manifold with negative and constant Ricci scalar.
Although there is no general relativistic limit in this theory, $\alpha$ can be positive indeed.
These results show that a wormhole can be constructed with ordinary matter in Einstein-Gauss-Bonnet gravity.

The purpose of the present paper is to investigate the $n(\ge 5)$-dimensional static Lorentzian wormholes in Einstein-Gauss-Bonnet gravity with ordinary matter respecting energy conditions.
For simplicity, the spacetime is supposed to have symmetries corresponding to the isometries of an $(n-2)$-dimensional maximally symmetric space, which is also assumed to be compact to make physical quantities finite. 
Our main results contain a part of the results obtained in~\cite{bk1992,dot2007} but are performed in a more simplified manner adopting the double-null coordinates.

The rest of the present paper is constituted as follows.
In the following section, a concise overview of Einstein-Gauss-Bonnet gravity, 
the definitions of some concepts used in the present paper, and basic equations are given. 
In section~III, our main results about the (non-)existence and the size of the wormhole throat, and the energy condition at the wormhole throat are presented by local analyses.
In section~IV, we construct exact static wormhole solutions with a dust fluid as explicit examples of wormhole solutions satisfying the energy conditions everywhere.
Concluding remarks and discussions including future prospects are summarized in section~V.

The conventions of the curvature tensors are 
$[\nabla _\rho ,\nabla_\sigma]V^\mu ={R^\mu }_{\nu\rho\sigma}V^\nu$ 
and $R_{\mu \nu }:={R^\rho }_{\mu \rho \nu }$.
The Minkowski metric is taken to be mostly the plus sign, and 
Roman indices run over all spacetime indices.

\section{Preliminaries}
We begin by a brief description of Einstein-Gauss-Bonnet gravity in the presence of a cosmological constant.
The action in $n (\geq 5)$-dimensional spacetime is given by
\begin{align}
\label{action}
S=\int\D ^nx\sqrt{-g}\biggl[\frac{1}{2\kappa_n^2}
(R-2\Lambda+\alpha{L}_{\rm GB}) \biggr]+S_{\rm matter},
\end{align}
where
$R$ and $\Lambda$ are the $n$-dimensional Ricci scalar and the cosmological constant, respectively. 
$S_{\rm matter}$ in Eq.~(\ref{action}) is the action for matter fields and $\kappa_n := \sqrt{8\pi G_n}$, where $G_n$ is the $n$-dimensional gravitational constant.
The Gauss-Bonnet term $L_{\rm GB}$ comprises the combination of the Ricci scalar, 
Ricci tensor $R_{\mu\nu}$, and Riemann tensor ${R^\mu}_{\nu\rho\sigma}$ as
\begin{align}
{L}_{\rm GB} := R^2-4R_{\mu\nu}R^{\mu\nu}+R_{\mu\nu\rho\sigma}R^{\mu\nu\rho\sigma}.
\end{align}
In four-dimensional spacetime, the Gauss-Bonnet term does not contribute to the field equations since it
becomes a total derivative.
$\alpha$ is the coupling constant of the Gauss-Bonnet term. 
This type of action is derived in the low-energy limit 
of heterotic string theory~\cite{Gross}.
In that case, $\alpha$ is regarded as the inverse string tension 
and positive-definite. 
However, we leave the sign of $\alpha$ unfixed 
in the present paper. 

The gravitational equation of the action (\ref{action}) is
\begin{align}
{G^\mu}_{\nu} +\alpha {H}^\mu_{~~\nu} 
+\Lambda \delta^\mu_{~~\nu}= 
\kappa_n^2 {T}^\mu_{~~\nu}, \label{beq}
\end{align}
where 
\begin{align}
{G}_{\mu\nu}&:= R_{\mu\nu}-{1\over 2}g_{\mu\nu}R,\\
{H}_{\mu\nu}&:= 2\Bigl[RR_{\mu\nu}-2R_{\mu\alpha}R^\alpha_{~\nu}-2R^{\alpha\beta}R_{\mu\alpha\nu\beta} \nonumber \\
&~~~~~~+R_{\mu}^{~\alpha\beta\gamma}R_{\nu\alpha\beta\gamma}\Bigr]
-{1\over 2}g_{\mu\nu}{L}_{\rm GB}
\end{align}
and ${T}^\mu_{~~\nu}$ is the energy-momentum tensor for matter fields obtained from $S_{\rm matter}$.
The field equation (\ref{beq}) contains up to the second derivatives of the metric.

Suppose the $n$-dimensional spacetime 
$({\ma M}^n, g_{\mu \nu })$ is a warped product of an 
$(n-2)$-dimensional constant curvature space $(K^{n-2}, \gamma _{ij})$ and a two-dimensional orbit spacetime $(M^2, g_{ab})$ under the isometry of $(K^{n-2}, \gamma _{ij})$. 
The line element may be written locally in the double-null coordinates as
\begin{align}
\D s^2 = -2e^{-f(u,v)}\D u\D v+r(u,v)^2 \gamma_{ij}\D z^i\D z^j. \label{metric}
\end{align}  
Null vectors $(\partial /\partial u)$ and $(\partial /\partial v)$ are taken to be future-pointing. 
Here $r$ is a scalar on $(M^2, g_{ab})$  with $r=0$ defining its boundary, and $\gamma_{ij}$ is the unit metric on $(K^{n-2}, \gamma _{ij})$ with its sectional curvature $k = \pm 1, 0$. 
We assume that $({\ma M}^n, g_{\mu \nu})$ is strongly causal, $(K^{n-2}, \gamma _{ij})$ is compact, and the metric $g_{\mu \nu}$ is at least $C^2$. 
It is worthwhile to mention that the null coordinates still have the rescaling freedoms of $u\to U(u)$ and $v\to V(v)$, leaving the metric (\ref{metric}) invariant.

Since the rank-two symmetric tensors on the maximally symmetric space
are proportional to the metric tensor,
the symmetry of the background spacetime determines the structure of
the energy momentum tensor $T_{\mu\nu}$ as
\begin{align}
T_{\mu\nu}\D x^\mu \D x^\nu =
&T_{uu}(u,v)\D u^2+2T_{uv}(u,v)\D u\D v \nonumber \\
&
+T_{vv}(u,v)\D v^2+p(u,v)r^2 \gamma_{ij}\D z^i\D z^j, \label{em}
\end{align}  
where $p(y)$ is a scalar function on $(M^2, g_{ab})$.
Then, the governing equation~(\ref{beq}) gives
\begin{widetext}
\begin{align}
&(r_{,uu}+f_{,u}r_{,u})\left[1+\frac{2{\tilde\alpha}}{r^2}
(k+2e^{f}r_{,u}r_{,v})\right]
=-\frac{\kappa_n^2}{n-2} r T_{uu}, \label{nullbasic1} \\
&(r_{,vv}+f_{,v}r_{,v})\left[1+\frac{2{\tilde\alpha}}{r^2}
(k+2e^{f}r_{,u}r_{,v})\right]
=-\frac{\kappa_n^2}{n-2} r T_{vv}, \label{nullbasic2} \\
&rr_{,uv}+(n-3)r_{,u}r_{,v}+\frac{n-3}{2}k e^{-f}+\frac{{\tilde\alpha}}{2r^2}
[(n-5)k^2e^{-f}+4rr_{,uv}
(k+2e^fr_{,u}r_{,v})+4(n-5)r_{,u}r_{,v}
(k+e^fr_{,u}r_{,v})] \nonumber \\
&~~~~~~-\frac{n-1}{2}{\tilde\Lambda}r^2e^{-f}
=\frac{\kappa_n^2}{n-2} r^2T_{uv}, \label{nullbasic3} \\
&r^2 f_{,uv}+2(n-3)r_{,u}r_{,v}
+k(n-3)e^{-f}-(n-4)rr_{,uv} \nonumber \\
&~~~~~~+\frac{2{\tilde\alpha}e^{-f}}{r^2}
\biggl[e^f(k+2e^fr_{,u}r_{,v})
\{r^2f_{,uv}-(n-8)rr_{,uv}\}
+2r^2e^{2f}(f_{,u}r_{,u} r_{,vv}
+f_{,v}r_{,v}r_{,uu}) \nonumber \\
&~~~~~~+(n-5)(k+2e^fr_{,u}r_{,v})^2
+2r^2e^{2f}\{r_{,uu}r_{,vv}
+f_{,u}f_{,v}r_{,u}r_{,v}
-(r_{,uv})^2\}\biggl] \nonumber \\
&~~~~~~=\kappa_n^2 r^2(T_{uv}+e^{-f}p), \label{nullbasic4}
\end{align}  
\end{widetext}
where ${\tilde \alpha} := (n-3)(n-4)\alpha$, 
${\tilde \Lambda} := 2\Lambda /[(n-1)(n-2)]$ and a comma denotes the partial differentiation.

The area expansions along two independent future-directed radial null vectors $(\partial/\partial v)$ and $(\partial/\partial u)$ are respectively defined as
\begin{align}
\theta_{+}&:=(n-2)r^{-1}r_{,v},\\
\theta_{-}&:=(n-2)r^{-1}r_{,u}.
\end{align}  
An invariant combination $e^f \theta_+\theta_-$ characterizes the marginal surface, as will be mentioned later in the present section.
On the other hand, the function $r$ has a geometrical meaning as an areal radius; the area of symmetric subspace is given by $A:=V^k_{n-2}r^{n-2}$, where $V_{n-2}^k$ is the area of the unit $(n-2)$-dimensional space of constant curvature.

The generalized Misner-Sharp quasi-local mass in Einstein-Gauss-Bonnet gravity in the presence of a cosmological constant~\cite{maeda2006b,mn2007} is defined by 
\begin{align}
m &:= \frac{(n-2)V_{n-2}^k}{2\kappa_n^2}
\biggl\{-{\tilde \Lambda}r^{n-1}
+r^{n-3}[k-(D r)^2] \nonumber \\
&~~~~~~+{\tilde \alpha}r^{n-5}[k-(Dr)^2]^2 \biggl\},
\end{align}  
where $D_a $ is a metric compatible linear connection on $(M^2, g_{ab})$
and $(Dr)^2:=g^{ab}(D_ar)(D_br)$.
In the double-null coordinates, it is expressed as
\begin{align}
\label{qlm}
m &= \frac{(n-2)V_{n-2}^k}{2\kappa_n^2}r^{n-3}
\biggl[-{\tilde \Lambda}r^2+\left(k+\frac{2}{(n-2)^2} r^2e^{f}
\theta_{+}\theta_{-}\right)\nonumber \\
&~~~~~~
+{\tilde \alpha}r^{-2}\left(k+\frac{2}{(n-2)^2} 
r^2e^{f}\theta_{+}\theta_{-}\right)^2\biggl].
\end{align}  
The properties of the above quantity such as the monotonicity or positivity were fully investigated in~\cite{mn2007} and it was shown to be a natural counterpart of the Misner-Sharp mass in four-dimensional spherically symmetric spacetimes without a cosmological constant~\cite{ms1964}.
From the equation above, we obtain
\begin{widetext}
\begin{align}
\label{trapping}
&\frac{2}{(n-2)^2} r^2 e^f\theta_{+}\theta_{-} = 
-k-\frac{r^2}{2{\tilde\alpha}}
\left(1\mp\sqrt{1+\frac{8\kappa_n^2{\tilde\alpha} m}
{(n-2)V^k_{n-2}r^{n-1}}+4{\tilde\alpha}{\tilde\Lambda}}\right).
\end{align}  
\end{widetext}

There are two families of solutions corresponding to the sign in front of the square root in Eq.~(\ref{trapping}).
We call the family having the minus (plus) sign 
the general relativistic (GR) branch (non-GR branch) solution.
Note that the GR-branch solutions have the general
relativistic limit as $\alpha \to 0$, but the 
non-GR-branch solutions do not.
Throughout this paper, the upper sign is used for the GR branch, 
i.e., the solution having the general relativistic limit.
Here we define a branch surface, where two branches of solutions degenerate.
\begin{dn}
\label{def:b-point}
A {\it branch surface} is an $(n-2)$-surface where inside the square root in Eq.~(\ref{trapping}) vanishes.
\end{dn}

Instead of specifying the matter fields, energy conditions
are imposed in our analysis.
The null energy condition for the matter field implies
\begin{align}
T_{uu}\ge 0,~~~T_{vv} \ge 0, \label{nec}
\end{align}
while the dominant energy condition implies
\begin{align}
T_{uu} \ge 0,~~T_{vv}\ge 0,~~T_{uv}\ge 0. \label{dec}
\end{align}  
The dominant energy condition assures that a causal observer measures the 
non-negative energy density, and the energy flux is 
a future-directed causal vector.

Here we recapitulate the local notions of spacetimes for later investigations.
\begin{dn}
\label{def:t-surface}
A {\it trapped (untrapped) surface} is an $(n-2)$-surface with $\theta_{+}\theta_{-}>(<)0$.
\end{dn}
\begin{dn}
\label{def:m-surface}
A {\it marginal surface} is an $(n-2)$-surface with $\theta_{+}\theta_{-}=0$.
\end{dn}
Without loss of generality, we set $\theta_{+}$ to be zero on a marginal surface.
Marginal surfaces are classified into several types depending on the sign of $\theta_-$ and $\theta_{+,u}$ there~\cite{hayward1994}. 
\begin{dn}
\label{def:4-msphere}
A marginal surface is {\it future} if $\theta_-<0$, {\it past} if
$\theta_{-}>0$, {\it bifurcating} if $\theta_-=0$, {\it outer} if
$\theta_{+,u}<0$, {\it inner} if $\theta_{+,u}>0$, and {\it
degenerate} if $\theta_{+,u}=0$.
\end{dn}

Now we give a definition of a wormhole throat by
\begin{dn}
\label{def:wormhole}
A wormhole throat is an $(n-2)$-surface with $r>0$ and 
\begin{eqnarray}
A_{,\mu}\zeta^\mu&=&0, \label{throatA} \\
(A_{,\mu}\zeta^\mu)_{,\nu}\zeta^\nu &>& 0, \label{throatB}
\end{eqnarray}  
where $A:=V^k_{n-2}r^{n-2}$ is the area of the $(n-2)$-surface and 
$\zeta^\mu(\partial/\partial x^\mu)=\zeta^u(\partial/\partial u)+\zeta^v(\partial/\partial v)$ 
is a spacelike vector, i.e., $\zeta^u\zeta^v<0$.
\end{dn}
The above definition means that an $(n-2)$-surface is said to be a wormhole 
throat if it has a positive minimum area on a spacelike hypersurface of constant time.
Equations~(\ref{throatA}) and (\ref{throatB}) are equivalent to 
\begin{eqnarray}
r_{,\mu}\zeta^\mu&=&0, \label{throat1} \\
(r_{,\mu}\zeta^\mu)_{,\nu}\zeta^\nu &>& 0. \label{throat2}
\end{eqnarray}  
By construction, our definition depends on time-slicing.

\section{Properties of static wormhole spacetimes}

In this section, we present our main results on the properties of static wormhole spacetimes.
Staticity is defined by the existence of a hypersurface-orthogonal
timelike Killing vector $\xi^\mu$. 
The Frobenius' integrability condition implies that there exist
scalar functions $t$ and $N$ such that $\xi_\mu=-N\nabla_\mu t$. 
Here we choose $N=e^{-f}$ and $t=(u+v)/\sqrt{2}$ without loss of generality,
and then we have $\xi^\mu(\partial/\partial x^\mu)=[(\partial/\partial u)+(\partial/\partial v)]/\sqrt{2}$.
This choice is always achieved by the rescaling freedom of the 
null coordinates. 
With this choice made, 
 the Killing equation reduces to
\begin{eqnarray}
r_{,u}+r_{,v}=0,\label{r-static}\\
f_{,u}+f_{,v}=0.
\end{eqnarray}  

Because we have $\xi_\mu\xi^\mu=N^2\nabla_\mu t\nabla^\mu t=-e^{-f}$ in the double-null coordinates (\ref{metric}), $e^{-f}$ is strictly positive, or equivalently $e^{f}$ is positive and finite, in static spacetimes.
This fact will be used implicitly in later calculations.

In static spacetimes, there exists a natural 
time-slicing $t={\rm constant}$ corresponding to the constant Killing time.
Respecting this symmetry, 
we naturally set the spacelike vector $\zeta^\mu$ in Definition~\ref{def:wormhole} such that 
$g_{\mu \nu}\xi^\mu \zeta^\nu=0$, from which we have
$\zeta^u=-\zeta^v$.
In the foregoing discussion, we will stick to this time-slicing to define a wormhole throat.

Equation~(\ref{r-static}) implies that a marginal surface in the static
spacetime is a bifurcating marginal surface independent of the field equations, i.e., the theories of gravity.
Here it should be emphasized that the marginal surface in the maximally 
extended Schwarzschild-Tangherlini-type spacetime is not a
counterexample against the
above statement, because the hypersurface-orthogonal Killing vector in that spacetime associated with the time translation at spacelike infinity becomes null at the marginal surface in that spacetime, so that the spacetime is not static there.
This is compatible to the fact that the ``wormhole throat'' 
in the Schwarzschild-Tangherlini spacetime in the isotropic coordinates is a coordinate singularity.
In other words, that spacetime, which reduces to the Einstein-Rosen
bridge in four dimensions, represents a static wormhole {\it without} a throat.

Indeed, the analysis in the static spacetime is drastically simplified thanks to the following lemma claiming that a wormhole throat in the static spacetime, in any theories of gravity, is necessarily a bifurcating marginal surface.
\begin{lm}
\label{pr:static2}
A wormhole throat in the static spacetime is a bifurcating marginal surface.
\end{lm}
\noindent 
{\it Proof}. 
Equation~(\ref{throat1}) with $\zeta^u=-\zeta^v$ implies
\begin{eqnarray}
r_{,u}-r_{,v}=0 \label{throat-static}
\end{eqnarray}  
at the wormhole throat.
From Eqs.~(\ref{r-static}) and (\ref{throat-static}), we obtain $\theta_+=\theta_-=0$ there.
\qed

\bigskip

In Einstein-Gauss-Bonnet gravity, the existence of a marginal surface 
and its location are highly restricted depending on the
signs of the sectional curvature $k$ and the Gauss-Bonnet coupling $\alpha$, as well
as the branches. 
This is in sharp contrast to the
general relativistic case. 
The following assertion is verified by direct calculations of Eq.~(\ref{trapping}).
\begin{lm}
\label{th:absenceTH}
Let $\alpha$ be positive. 
Then, an $(n-2)$-surface is necessarily untrapped, and marginal surfaces are 
absent in the non-GR-branch solutions for $k=0$ and $1$.
In the GR-branch (non-GR-branch) solutions for $k=-1$ 
with $r^2<(>)2{\tilde\alpha}$, an $(n-2)$-surface is necessarily trapped 
(untrapped), and marginal surfaces are absent. 
Let $\alpha$ be negative. 
Then, an $(n-2)$-surface is necessarily trapped, and marginal surfaces are 
absent in the non-GR-branch solutions for $k=0$ and $-1$.
In the GR-branch (non-GR-branch) solutions for $k=1$ 
with $r^2<(>)2|{\tilde\alpha}|$, an $(n-2)$-surface is necessarily untrapped 
(trapped), and marginal surfaces are absent.
\end{lm}

\begin{center}
\begin{table}[h]
\caption{\label{table:m-surface} The allowed region of a marginal surface by Lemma~\ref{th:absenceTH} in Einstein-Gauss-Bonnet gravity.}
\begin{tabular}{l@{\quad}|@{\quad}c@{\qquad}c@{\quad}|@{\quad}c@{\qquad}c}
\hline \hline
&\multicolumn{2}{c|@{\quad}}{GR branch} &
\multicolumn{2}{c}{non-GR branch} \\
 & $\alpha>0$ & $\alpha<0$ & $\alpha>0$ & $\alpha<0$ \\\hline
$k=1$ & Any $r$ &  $r^2 \ge 2|{\tilde\alpha}|$ &  None & $r^2 \le 2|{\tilde\alpha}|$ \\\hline
$k=0$ & Any $r$ & Any $r$ & None & None \\\hline
$k=-1$ & $r^2\ge2{\tilde\alpha}$ & Any $r$ & $r^2 \le 2{\tilde\alpha}$ & None \\ \hline
\hline
\end{tabular}
\end{table} 
\end{center}
The allowed region of a marginal surface by Lemma~\ref{th:absenceTH} is summarized in Table~\ref{table:m-surface}.
Combining Lemmas~\ref{pr:static2} and \ref{th:absenceTH}, 
we achieve the following non-existence theorem for the wormhole throat
in the static spacetime.
\begin{Prop}
\label{pr:static2}
There is no wormhole throat in the non-GR-branch static 
solutions for $k\alpha \ge 0$ and in the GR-branch (non-GR-branch) 
static solutions for $k\alpha<0$ with $r^2<(>)2|{\tilde\alpha}|$.
\end{Prop}

By the proposition above, the existence 
of a wormhole throat is then restricted to the following three possibilities: 
(i) the GR-branch solutions with $k\alpha \ge 0$, 
(ii) the GR-branch solutions with $k\alpha<0$ and a sufficiently thick
wormhole throat, 
and (iii) the non-GR-branch solutions with $k\alpha<0$ and a sufficiently
thin wormhole throat.
But are all of these three cases equally realized in physically reasonable 
circumstances? 
The next three propositions concerning the energy condition at the
wormhole throat give a partial answer to this question.

\begin{Prop}
\label{pr:static3}
The null energy condition is violated at the wormhole throat with $r_{\rm th}^2 \ne -2k{\tilde\alpha}$ in the static spacetime in the GR branch, where $r_{\rm th}$ is the areal radius of the wormhole throat.
\end{Prop}
\begin{Prop}
\label{pr:static4}
In static spacetimes in the non-GR branch,
$T_{uu}>0$ and $T_{vv}>0$ are satisfied at the wormhole throat with $r_{\rm th}^2 \ne -2k{\tilde\alpha}$.
In the case of $\Lambda \le 0$, $T_{uv}\ge 0$ is also satisfied there if $(n-4)(n-5)|{\alpha}| \le r_{\rm th}^2 < 2(n-3)(n-4)|{\alpha}|$ holds for $k=1$ and $\alpha<0$ or if $r_{\rm th}^2 \le (n-4)(n-5){\alpha}$ holds for $k=-1$ and $\alpha>0$.
\end{Prop}
\noindent 
{\it Proof}. 
Differentiating Eq.~(\ref{r-static}) with respect to $u$ and $v$, we respectively obtain 
\begin{eqnarray}
r_{,vu}+r_{,uu}=0 \label{rvu=-rvv}
\end{eqnarray}  
and 
\begin{eqnarray}
r_{,uv}+r_{,vv}=0, \label{ruv=-ruu}
\end{eqnarray}  
which give
\begin{eqnarray}
r_{,uu}=r_{,vv}=-r_{,vu}. \label{ruu=-rvv}
\end{eqnarray}
Equations~(\ref{throat2}) and (\ref{ruu=-rvv}), together with Lemma~\ref{pr:static2}, give 
\begin{eqnarray}
r_{,uu}(\zeta^u-\zeta^v)^2>0 \label{ruu^2}
\end{eqnarray}
at the wormhole throat.
Because $\zeta^u=-\zeta^v$ is non-zero, we have $r_{,uu}>0$ and consequently $r_{,vv}>0$ and $r_{,uv}<0$ there by Eq.~(\ref{ruu=-rvv}).
On the other hand, Eqs.~(\ref{nullbasic1}), (\ref{nullbasic2}) and (\ref{nullbasic3}), together with Lemma~\ref{pr:static2}, imply that 
\begin{eqnarray}
&&r_{,uu}\biggl(1+\frac{2k{\tilde\alpha}}{r^2}\biggl)=-\frac{\kappa_n^2}{(n-2)}r T_{uu}, \label{ruu} \\
&&r_{,vv}\biggl(1+\frac{2k{\tilde\alpha}}{r^2}\biggl)=-\frac{\kappa_n^2}{(n-2)}r T_{vv}, \label{rvv} \\
&&rr_{,uv}\biggl(1+\frac{2k{\tilde\alpha}}{r^2}\biggl)+\frac{n-3}{2}k e^{-f}\biggl[1+\frac{(n-5)k{\tilde\alpha}}{(n-3)r^2}\biggl] \nonumber \\
&&~~~~~~-\frac{n-1}{2}{\tilde\Lambda}r^2e^{-f}
=\frac{\kappa_n^2}{n-2} r^2T_{uv} \label{ruv} 
\end{eqnarray}  
hold at the wormhole throat.

Thus, by Eqs.~(\ref{ruu}) and (\ref{rvv}) together with Proposition~\ref{pr:static2}, $T_{uu}<(>)0$ and $T_{vv}<(>)0$ are satisfied at the wormhole throat in the GR (non-GR) branch with $r_{\rm th}^2 \ne -2k{\tilde\alpha}$, which proves Proposition~\ref{pr:static3} and the first part of Proposition~\ref{pr:static4}.

By Proposition~\ref{pr:static2}, a wormhole throat exists in the static non-GR-branch solutions only for $k\alpha<0$ with $r_{\rm th}^2 \le -2k{\tilde\alpha}$.
Then, the first term in the left-hand-side of Eq.~(\ref{ruv}) is non-negative at the wormhole throat in the non-GR branch, while the last term is also non-negative for $\Lambda \le 0$.
Therefore, $T_{uv}\ge 0$ holds at the wormhole throat if $k[r_{\rm th}^2+(n-4)(n-5)k{\alpha}] \ge 0$ is satisfied.
Because $-(n-4)(n-5)k{\alpha} < -2k{\tilde\alpha}$ is satisfied for $k\alpha<0$, the above condition reduces to $-(n-4)(n-5)k{\alpha} \le r_{\rm th}^2 \le -2k{\tilde\alpha}$ for $k=1$ and $\alpha<0$, while it reduces to $r_{\rm th}^2 \le -(n-4)(n-5)k{\alpha}$ for $k=-1$ and $\alpha>0$.
This completes the proof of Proposition~\ref{pr:static4}.
\qed

\bigskip

In the two propositions above, we only considered the wormhole throat with $r_{\rm th}^2 \ne -2k{\tilde\alpha}$. 
A wormhole throat with $r_{\rm th}^2=-2k{\tilde\alpha}$ corresponds to a branch surface, so that two branches cannot be distinguished locally from each other.
This case is rather special and should be treated separately.
\begin{Prop}
\label{pr:special}
$T_{uu}=T_{vv}=0$ is satisfied at the static wormhole throat with $r_{\rm th}^2=-2k{\tilde\alpha}$.
Also, $T_{uv}>(<)0$ and $T_{uv}=0$ are satisfied there for ${\alpha}(1+4{\tilde\alpha}{\tilde\Lambda})<(>)0$ and $1+4{\tilde\alpha}{\tilde\Lambda}=0$, respectively.
\end{Prop}
\noindent 
{\it Proof}. 
Equations~(\ref{ruu}) and (\ref{rvv}) give $T_{uu}=T_{vv}=0$ at the wormhole throat with $r_{\rm th}^2=-2k{\tilde\alpha}$, while Eq.~(\ref{ruv}) gives
\begin{eqnarray}
&&\frac{n-1}{4}e^{-f}(1+4{\tilde\alpha}{\tilde\Lambda})=-\frac{2\kappa_n^2{\tilde\alpha}}{n-2}T_{uv} 
\end{eqnarray}  
there, where $k \ne 0$ was used because of $r_{\rm th}>0$.
Thus, we have $T_{uv}>(<)0$ and $T_{uv}=0$ at the wormhole throat for ${\alpha}(1+4{\tilde\alpha}{\tilde\Lambda})<(>)0$ and $1+4{\tilde\alpha}{\tilde\Lambda}=0$, respectively. 
\qed

The results obtained up to this point are summarized in Table~\ref{table:wormhole}.
Propositions~\ref{pr:static3} and \ref{pr:special} imply that the wormhole throat in a static solution respecting the energy conditions in the GR branch is necessarily a branch surface and then ${\alpha}(1+4{\tilde\alpha}{\tilde\Lambda}) \le 0$ is required.
Although a branch surface is a curvature singularity if $T_{uu} \ne 0$ or $T_{vv} \ne 0$ holds there~\cite{nm2007}, it could be regular if we have $T_{uu}=T_{vv}=0$ there. 
Of course, even if such a solution is successfully constructed, its general relativistic limit does not represent a wormhole since we have $r_{\rm th} \to 0$ for $\alpha \to 0$.

\begin{widetext}
\begin{center}
\begin{table}[h]
\caption{\label{table:wormhole} Properties of the wormhole throat in the static spacetime in Einstein-Gauss-Bonnet gravity. NEC stands for the null energy condition. See Proposition~\ref{pr:special} for the case where the wormhole throat coincides with a branch surface.}
\begin{tabular}{l@{\quad}|@{\quad}c@{\qquad}c@{\quad}|@{\quad}c@{\qquad}c}
\hline \hline
&\multicolumn{2}{c|@{\quad}}{GR branch} &
\multicolumn{2}{c}{non-GR branch} \\
 & $\alpha>0$ & $\alpha<0$ & $\alpha>0$ & $\alpha<0$ \\\hline
$k=1$ & NEC violation &  NEC violation &  Absent & See Proposition~\ref{pr:static4} \\\hline
$k=0$ & NEC violation & NEC violation & Absent & Absent \\\hline
$k=-1$ & NEC violation & NEC violation & See Proposition~\ref{pr:static4} & Absent \\ \hline
\hline
\end{tabular}
\end{table} 
\end{center}
\end{widetext}

On the other hand,
in the non-GR-branch solutions,
the inequalities~(\ref{dec}) hold at the wormhole throat for $k\alpha<0$ and $\Lambda \le 0$ when the radius of the throat satisfies the inequality in Proposition~\ref{pr:static4}.
Here we emphasize that the inequalities~(\ref{dec}) are not a sufficient condition for the dominant energy condition.
Actually, the inequalities~(\ref{dec}) are identical to the dominant energy condition only for a radial null vector.
Even under the inequalities~(\ref{dec}), the null energy condition can
be violated for a non-radial null vector 
provided the function $p$ in the energy momentum tensor~(\ref{em}) is negative and sufficiently large.
This is apprehensible by writing down the null energy condition for
a generic null vector $k^\mu$ as
\begin{align} 
T_{\mu\nu}k^\mu k^\nu&=T_{uu}(k^u)^{2}+T_{vv}(k^u)^2+2T_{uv}k^uk^v+pr^2\gamma_{ij}k^ik^j, \nonumber \\
&=T_{uu}(k^u)^2+T_{vv}(k^v)^2+2(T_{uv}+pe^{-f})k^{u}k^{v}, \nonumber \\
&\ge 0,
\end{align} 
where $k^\mu k_\mu=0$ was used at the second equality.
Thus, for a matter field with $p=0$ such as vacuum or a dust fluid, the conditions in Proposition~\ref{pr:static4} are surely sufficient for the dominant energy condition to hold at the wormhole throat.

Recently, Dotti, Oliva and Troncoso obtained a wormhole solution in the five-dimensional vacuum case with $1+4{\tilde\alpha}{\tilde\Lambda}=0$, in which the three-dimensional submanifold has a negative and constant Ricci scalar~\cite{dot2007}.
Their result includes the case with a three-dimensional negative
constant curvature.
This special tuning between the coupling constants allows the theory to have a unique maximally symmetric solution~\cite{ctz2000} and yields the Chern-Simons gravity in five dimensions, which is the lowest number of dimensions in which the Gauss-Bonnet term becomes nontrivial~\cite{Banados:1993ur}.
At first glance, solutions are strongly restricted by this special relation among all the solutions with arbitrary coupling constants. 
However, this turns out not to be the case for wormhole solutions.
Indeed, as a corollary of the next proposition, it is shown that this relation, $1+4{\tilde\alpha}{\tilde\Lambda}=0$, 
is a necessary condition for vacuum static wormhole solutions in Einstein-Gauss-Bonnet gravity.

\begin{Prop}
\label{pr:static5}
If $T_{uu}=T_{vv}=T_{uv}=0$ is satisfied at the wormhole throat, 
$1+4{\tilde\alpha}{\tilde\Lambda}=0$ holds and the radius 
of the wormhole throat is given by $r_{\rm th}^2=-2k{\tilde \alpha}$.
\end{Prop}
\noindent 
{\it Proof}. 
Equations~(\ref{ruu}) and (\ref{rvv}) with the fact that $r_{,uu}>0$ and $r_{,vv}>0$ give $r_{\rm th}^2=-2k{\tilde \alpha}$ and hence $k \ne 0$.
Then, the evaluation of Eq.~(\ref{ruv}) at the wormhole throat gives $1+4{\tilde\alpha}{\tilde\Lambda}=0$.
\qed

\bigskip

Thus, in vacua, $k=-(+)1$ and $\Lambda<(>)0$ are required for positive (negative) $\alpha$.
Furthermore, it follows from Eq.~(\ref{qlm}) that $m=0$ holds at the wormhole throat in the vacuum case.
Because $m$ is constant in the vacuum case~\cite{mn2007}, we obtain $m \equiv 0$ in the whole spacetime, hence any $r$ is a branch surface.

The vacuum solutions with $1+4{\tilde\alpha}{\tilde\Lambda}=0$ are
completely classified into three classes, namely the Nariai-type
solution, the generalized Boulware-Deser-Wheeler solution, and the class I solution~\cite{birkhoffGB,mn2007}. 
(Proposition 1 in~\cite{mn2007} also holds for negative $\alpha$.) 
We note that the class I solution is not necessarily static.
The Nariai-type solution is not a wormhole spacetime, because the areal radius is constant~\cite{Lorenz-Petzold1987a,md2007}.
Since the hypersurface-orthogonal Killing vector becomes a zero vector 
at the bifurcating marginal surface in the generalized 
Boulware-Deser-Wheeler solution~\cite{EGBBH,tm2005}, 
it does not contain a wormhole throat as in the Schwarzschild-Tangherlini case.
The static metric of the class I solution with $k \ne 0$ can be written as
\begin{align}
\D s^2=2 |{\tilde\alpha}|\biggl[-e^{2\phi(\rho)}\D t^2 +\D \rho^2+\cosh^2(\sqrt{-k}\rho) \gamma _{ij}\D z^i\D z^j\biggl],
\label{eq:staticmetric}
\end{align}
where $\phi(\rho)$  is an arbitrary function of $\rho$.

The metric~(\ref{eq:staticmetric}) with $k=1$ and $\alpha<0$ does not represent a wormhole independent of $\phi(\rho)$ because the conditions~(\ref{throatA}) and (\ref{throatB}) do not hold for $r>0$.
For a function $\phi(\rho)$ leaving $\rho=0$ non-singular, on the other hand, the metric~(\ref{eq:staticmetric}) with $k=-1$ and $\alpha>0$ includes a wormhole throat at $\rho=0$, of which areal radius is given by $r_{\rm th}=\sqrt{2{\tilde \alpha}}$, compatible to Proposition~\ref{pr:static5}.

In this section, we have seen the static wormhole throat may exist under the null energy condition in Einstein-Gauss-Bonnet gravity.
While a wormhole throat in the GR branch must be a branch surface with a fixed radius, its size is less restricted in the non-GR branch.
The origin of the antithetical and pathological behaviors in the non-GR branch is presented as follows.

Using Eqs.~(\ref{nullbasic1})--(\ref{nullbasic4}) and (\ref{trapping}) together with the expressions of the Ricci tensors, we obtain
\begin{align}
\pm R_{\mu\nu}k^\mu k^\nu\sqrt{1+\frac{8\kappa_n^2{\tilde\alpha} m}
{(n-2)V^k_{n-2}r^{n-1}}+4{\tilde\alpha}{\tilde\Lambda}}
=\kappa _n^2T_{\mu\nu}k^\mu k^\nu
\label{Rvv}
\end{align}
for a radial null vector $k^\mu$, where 
$k^\mu(\partial/\partial x^\mu)=k^u(\partial/\partial u)$
 or $k^v(\partial/\partial v)$. (See Lemma~2 in~\cite{nm2007} for more details.)
We note that a branch surface is a degenerate point in Eq.~(\ref{Rvv}). 
Equation~(\ref{Rvv}) shows that the null convergence condition 
$R_{\mu\nu}k^\mu k^\nu \ge 0$ fails in the non-GR branch 
if the null energy condition is {\it strictly} satisfied $T_{\mu\nu}k^\mu k^\nu > 0$.
It signals that solutions in the non-GR branch behave badly 
under the null energy condition, since properties of the geometry are
determined not by energy conditions but by the convergence condition,
as seen in the Raychaudhuri equation. In the non-GR-branch solution, 
gravity effectively acts repulsively for
the positive energy particles. 

In closing this section, we make a short comment on the definition of a wormhole throat.
We have hitherto proceeded by considering that the vector 
 $\zeta^\mu$ in Definition~\ref{def:wormhole} as is a spatial vector orthogonal to the timelike Killing vector.
An alternative way to define a wormhole throat is to set $\zeta^\mu$ as a radial null vector, i.e., a wormhole throat is defined on a null hypersurface~\cite{hv1997,hayward1999}.
Then, from Eqs.~(\ref{throat1}) and (\ref{throat2}), 
the wormhole throat is given by $r_{,v}=0$ and $r_{,vv}>0$ or $r_{,u}=0$ and $r_{,uu}>0$.
On the other hand, Eqs.~(\ref{r-static}) and (\ref{ruu=-rvv}) 
respectively also give $r_{,u}=r_{,v}=0$ and $r_{,uu}=r_{,vv}>0$ at the wormhole throat in this case.
As a result, the wormhole throat coincides with that defined on a
 spacelike hypersurface with the constant Killing time, and all the propositions in the present paper
 remain valid in this case.

\section{Exact wormhole solutions respecting energy conditions}
Proposition~\ref{pr:static4} tells us that, in the non-GR branch, the energy conditions are
satisfied at the wormhole throat for $k\alpha<0$ in the case without tangential pressure.
However, it does not ensure the energy condition respected in the whole spacetime.
Bhawal and Kar have reported that even when the weak energy condition is
respected at the wormhole throat for $k=1$ and $\alpha<0$, 
it is impossible to make a $C^2$ wormhole solution in which the energy condition is satisfied everywhere~\cite{bk1992}.
However, their discussion is based on the positivity of the quantity $N-P$ in~\cite{bk1992}, which is shown only at the wormhole throat and seems not to be so valid in the whole spacetime.
In this section, although we do not give a counterexample to their claim, we show that their result cannot be extended to the case with $k=-1$ and $\alpha>0$ by constructing exact static wormhole solutions with a dust fluid respecting the energy conditions everywhere.

The energy-momentum tensor of a dust fluid is
\begin{eqnarray}
{T}_{\mu\nu}=\mu u_{\mu}u_{\nu},
\end{eqnarray}
where $u^\mu$ and $\mu$ are the $n$-velocity of the fluid element and energy density, respectively.
(See~\cite{maeda2006b} for the basic equations in the comoving coordinates.)
It is shown that the synchronous comoving coordinates are possible 
in the dust case even without the staticity assumption, i.e., 
the lapse function can be set to unity~\cite{maeda2006b}.

Now we focus on the static solution and adopt the proper length 
as a radial coordinate without loss of generality.
Then we have
\begin{align}
\label{met}
\D s^2&=-\D t^2+\D \rho^2+r(\rho)^2\gamma_{ij}\D z^i \D z^j,\\
u^\mu\frac{\partial}{\partial x^\mu}&=\frac{\partial}{\partial t}.
\end{align}  
The $(\rho,\rho)$ component of the field equation~(\ref{beq}) gives
\begin{align}
0&=(n-5){\tilde\alpha}(r')^{4}-[(n-3)r^2+2(n-5)k{\tilde\alpha}](r')^{2} \nonumber \\
&~~+k[(n-3)r^2+(n-5)k{\tilde\alpha}]-(n-1){\tilde\Lambda} r^4, \label{beqdust}
\end{align}  
where a prime denotes the derivative with respect to $\rho$.

For $n=5$, the above equation reduces to 
\begin{align}
0=3(r')^{2}-3k+\Lambda r^2. 
\end{align}  
For $k\Lambda \le 0$, $r'$ cannot be zero, so that there is no wormhole throat in this case.
For $k\Lambda>0$, the general solution is given by 
\begin{align}
r=\sqrt{\frac{3}{|\Lambda|}}\cosh\biggl(\sqrt{\frac{-k|\Lambda|}{3}}\rho\biggl),
\end{align}
where an integration constant was set to zero without loss of generality by the coordinate transformation of $\rho$.
The solution with $k=1$ and $\Lambda>0$ is discarded since it does not meet the requirement (\ref{throat2}) of having a wormhole throat.
On the other hand, the solution with $k=-1$ and $\Lambda<0$ represents a wormhole.
The energy density of the dust, given from the $(t,t)$ component of the field equation, and the quasi-local mass of this solution in the latter case are obtained by 
\begin{align}
\mu&=\frac{\Lambda(3+4\alpha\Lambda)}{3\kappa_5^2},\\
m&=\frac{3(3+4\alpha\Lambda)V_{3}^{-1}}{4\Lambda\kappa_5^{2}}\cosh^{4}\biggl(\sqrt{-\frac{\Lambda}{3}}\rho\biggl).
\end{align}

However, it is shown that this solution has non-positive energy density and cannot be an example of the non-vacuum wormhole solutions respecting the energy conditions.
Calculating Eq.~(\ref{trapping}), we find that this solution belongs to the GR-branch for $3+4\alpha\Lambda \ne 0$.
Then, by Lemmas~\ref{pr:static2} and \ref{th:absenceTH}, the areal radius of the wormhole throat for $n=5$ with $k=-1$ and $\alpha>0$ in the GR branch satisfies $r^2 \ge 4\alpha$.
For the above solution, this condition gives $3+4\alpha\Lambda \ge 0$, and therefore the energy density is non-positive.
For $3+4\alpha\Lambda=0$, the solution reduces to the vacuum class I solution (\ref{eq:staticmetric}) with $\phi \equiv 0$.

In the case of $n \ge 6$, on the other hand, Eq.~(\ref{beqdust}) gives
\begin{widetext}
\begin{align}
{r'}^{2}=\frac{(n-3)r^2+2(n-5)k{\tilde\alpha}\mp r^2\sqrt{(n-3)^2+4(n-1)(n-5){\tilde\alpha}{\tilde\Lambda}}}{2(n-5){\tilde\alpha}}.
\end{align}  
\end{widetext}
Hereafter we set $\Lambda=0$ for simplicity.
Then, the GR-branch solution cannot be a wormhole solution, because $r'$
is constant.
Turning to the non-GR branch, we have
\begin{align}
{r'}^{2}=\frac{(n-3)r^2+(n-5)k{\tilde\alpha}}{(n-5){\tilde\alpha}}.\label{beqdust2}
\end{align}  
The wormhole solution arises only if $k\alpha < 0$ by Proposition~\ref{pr:static2}. 
The general solution of Eq.~(\ref{beqdust2}) for $k\alpha <0$ is given by 
\begin{align}
r=\sqrt{-(n-4)(n-5)k\alpha}\cosh\biggl[\frac{\sqrt{-k}
\rho}{\sqrt{(n-4)(n-5)|\alpha|}}\biggl],
\end{align}   
which belongs to the non-GR branch.
The energy density and the quasi-local mass are obtained as
\begin{align}
\mu&=\frac{(n-1)(n-2)}{(n-4)(n-5)^2\kappa _n^2\alpha},\\
m&=-\frac{(n-2)kV_{n-2}^k[-(n-4)(n-5)\alpha k]^{(n-3)/2}}{(n-5)\kappa_n^{2}} \nonumber \\
&~~~~~~~~~~~\times \cosh^{n-1}\biggl[\frac{\sqrt{-k}\rho}{\sqrt{(n-4)(n-5)|\alpha|}}\biggl].
\end{align}
The solution for $k=1$ and $\alpha<0$ does not represent a wormhole spacetime.
On the other hand, the solution with $k=-1$ and $\alpha>0$ represents a wormhole respecting the energy conditions everywhere.
The radius of the wormhole throat is $r_{\rm th}=\sqrt{-(n-4)(n-5)k\alpha}$, which is consistent with Lemmas~\ref{pr:static2} and \ref{th:absenceTH} and coincides with the upper bound in Proposition~\ref{pr:static4}.
This solution shows that the claim by Bhawal and Kar is not extended to the case with $k=-1$ and $\alpha>0$.

\section{Summary and discussion}
In this paper, we investigated properties of $n(\ge 5)$-dimensional static wormhole spacetimes in Einstein-Gauss-Bonnet gravity. 
We supposed that the metric is at least $C^2$ and the spacetime has symmetries corresponding to the isometries of an $(n-2)$-dimensional constant curvature space.
A wormhole throat is defined by an $(n-2)$-surface with a positive minimum area on a spacelike hypersurface orthogonal to the timelike Killing vector.
The system with $k=1$ was previously studied by Bhawal and Kar in a different set of coordinates~\cite{bk1992}.
We generalized their analysis to the case with general $k$ and in the presence of a cosmological constant in a more simplified manner adopting the double-null coordinates.

Solutions are classified into two types, namely the GR and non-GR branches, depending on the existence or absence of the general relativistic limit $\alpha \to 0$.
In the GR branch, we showed that a static wormhole throat respecting the energy conditions necessarily coincides with a branch surface, otherwise the null energy condition is violated there.
In the non-GR branch, the absence of wormhole solutions was shown for $k\alpha \ge 0$.
In the non-GR branch with $k\alpha<0$ and $\Lambda \le 0$, we showed that the dominant energy condition holds at the wormhole throat if the matter field has zero tangential pressure and the areal radius of the throat satisfies some inequality.
Especially in the vacuum case, a special relation between the coupling constants $1+4{\tilde\alpha}{\tilde\Lambda}=0$, which yields Chern-Simons gravity in five dimensions, is shown to be a necessary condition for static wormhole spacetimes.
Then, the areal radius of a wormhole throat is fixed by $r_{\rm th}^2=-2k{\tilde \alpha}$, which is consistent with the result of Dotti, Oliva and Troncoso~\cite{dot2007}.

The analyses above are performed purely locally at the wormhole throat, 
and therefore it is not trivial whether a wormhole solution is possible with matter respecting the energy conditions everywhere.
We showed that such a solution is possible in the case of $n \ge 6$, $k=-1$ and
$\alpha>0$ by explicitly constructing exact solutions with a dust fluid.
Bhawal and Kar claimed that such a wormhole solution is impossible in the case of $k=1$ and $\alpha<0$~\cite{bk1992}.
Although their explanation is based on a non-trivial assumption which seems to be valid only at the wormhole throat, 
there is no counterexample against their result up to now as
far as the authors know.

It should be emphasized that the differentiability of the metric 
is crucially related to existence of the wormhole solution respecting the energy conditions.
In the present paper, the metric function is assumed to be at least $C^{2}$.
If we weakens this assumption to $C^0$, it has been shown in five dimensions that a vacuum wormhole solution is possible by gluing two Boulware-Deser-Wheeler solutions with a thin-shell respecting the energy condition~\cite{gw2007,rs2007,gggw2007}.
In that case, a special relation between $\alpha$ and $\Lambda$ 
is not necessary and the size of the wormhole throat can be arbitrary in contrast to the $C^2$ case.
(See Proposition~\ref{pr:static5}.)
In the present paper, we gave the first example of static and smooth Lorentzian exact wormhole solutions respecting energy conditions in the non-vacuum case in Einstein-Gauss-Bonnet gravity.

In the present paper, unfortunately, we have not obtained rigorous results in the non-GR branch with non-vanishing tangential pressure.
The main difficulty in this regard is to control the behavior of the metric function $f$ and its derivatives at the wormhole throat.
However, even in that case, the null energy condition can hold for a radial null vector, which signals the existence of wormhole solutions with ordinary matter.
Such an antithetical and pathological behavior in the non-GR-branch solutions can be understood by considering the relation between the focusing condition in the Raychaudhuri equation and the null energy condition~\cite{nm2007}.
For the further progress in this direction, wormhole solutions with a scalar field should be the focus of future research as the simplest and important matter field with non-vanishing tangential pressure in the higher-dimensional context.

Lastly, independent of the subject for investigation, an ambitious problem worth trying to solve is to distinguish two branches of solutions without using spacetime symmetry.
Together with the results in~\cite{nm2007}, our results indicate that the eccentric behaviors appear only in the non-GR-branch solutions, while the properties of the GR-branch solutions are quite similar to those in general relativity.
Then, it is natural to guess that this also holds even without spacetime symmetry.
If we can distinguish two branches in generic spacetimes, it could allow us to extend the strong results in general relativity to Einstein-Gauss-Bonnet gravity in the GR branch.

\section*{Acknowledgments}
The authors would like to thank B.J.~Carr, G.~Dotti, C.~Garraffo, G.~Giribet, T.~Harada, J.~Oliva, R.~Troncoso and S.~Willison for discussions and comments. 
MN would like to thank Kei-ichi Maeda for continuous encouragement. 
HM was supported by Fondecyt grant 1071125 and the Grant-in-Aid for Scientific Research Fund of the Ministry of
Education, Culture, Sports, Science and Technology, Japan (Young Scientists (B) 18740162).
The Centro de Estudios Cient\'{\i}ficos (CECS) is funded by the Chilean
Government through the Millennium Science Initiative and the Centers of
Excellence Base Financing Program of Conicyt. CECS is also supported by a
group of private companies which at present includes Antofagasta Minerals,
Arauco, Empresas CMPC, Indura, Naviera Ultragas and Telef\'{o}nica del Sur.
MN was partially supported by JSPS.


\end{document}